# Reconfigurable Multiplier Architecture based on Memristor-CMOS with Higher Flexibility

Seungbum Baek
College of Electrical and Computer Engineering
Chungbuk National University
Cheongju, South Korea
sbbaek@cbnu.ac.kr

*Abstract*—Multiplication is an indispensable operation in most of digital signal processing systems. Recently, many systems need to execute different types of algorithms on a multiplier. Therefore, it needs complicated computation and large area occupation. In this regard a fixed multiplier is inefficient and the development of a reconfigurable multiplier becomes increasingly important. The advent of memristor-CMOS hybrid circuits provides an opportunity for reducing area occupation. This paper introduces memristor-CMOS based reconfigurable multiplier which provides flexible multiplication according to various bit-width. Performance of the proposed multiplier is estimated with some applications and comparison with conventional multipliers, using memristor SPICE model and proprietary 180-nm CMOS process.

I. INTRODUCTION

Multiplication is an essential component in most of systems which perform digital signal processing. It is fully utilized by telecommunication systems and digital systems used by many people such as camera and printer. Therefore these days in order to process digital signal faster and with lower power dissipation, research in efficient multiplier has been much progressed. In digital signal processing the multiply-accumulate operation is a generic step that computes the product of two numbers and adds it to an accumulator usually described in the context of multiplier-accumulator. As multipliers are far slower than other components and requires huge area in a system, multiplier which features short propagation delay, low power consumption, and small area may decide the overall performance of a system.

Memristors have been widely used as reconfigurable analog multiplication blocks in crossbar arrays over the past half a decade [1-4]. However, sub-nanoampere read currents are required to surpass conventional computing efficiency which has proven difficult in learning [5]. Another issue is the variance associated with analog multiplication. While analog multipliers are of major use for near-sensor systems (by allowing reduction of AD conversions and associated speed-up) almost all commercial computing systems process in the digital domain. Success in analog processing systems would be transformative for near-sensor computing and in-memory processing, but it is equally important to develop digital computing systems as these will be immediately pervasive in modern systems. Recently, plenty of systems need to compute various mathematical algorithms which have variable input data size on a multiplier. Therefore it needs complex operation and huge area occupation. In this regard fixed multiplier is grossly inefficient in terms of processing speed and power consumption. However, most existing multipliers employ static sizes for input data and have been focused on improving operation speed. Consequently researching improved multiplier which has enhanced flexibility without falling-off in processing speed is needed [6].

Researchers have proposed reconfigurable or scalable multipliers. An approach in the design of a scalable multiplier is presented in [7] whereby propagation of carry and sum signals are controlled by incorporation of multiplexers within the multiplier. Madhuria presented an alternative architecture for integer and Galois field multiplication with controlled carry input [8]. In this architecture, the multiplier performs an integer multiplication when the carry input is activated, and it carries out Galois field multiplication when the carry input is deactivated. Sjalander introduced a multiplier which could compute twin precision multiplication [9]. The multiplier takes 3-input AND gate of which the one of input would act as an on-off switch for each of partial products. Nevertheless, precision of the multiplier is only N/2-bit or N-bit in an N-bit multiplier. And Ahmed proposed a twin precision multiplier with modified 2-D bypass logic [10]. The multiplier skips full adder cells in the columns and rows of zero bits.

Reconfigurable multiplier architectures described thus far are adequate techniques to improve system performance by reducing power dissipation and increasing design flexibility. However, there still remains the complexity associated with silicon area utilization by such architectures due to the additional embedded control elements. With the advent of memristor-CMOS process that combines nano-scale memristive devices [11] and CMOS processing, it becomes possible to reduce utilization of silicon area thus providing a promising option in the design of large-scale memories [12] and as reconfigurable and scalable multiplier arrays [13-14].

We propose the reconfigurable multiplier architecture based on a memristor-CMOS technology [14] to reduce the area occupation and increase flexibility for various applications according to the size of input data multiplier and multiplicand. This paper consists following orders. First of all, we introduce the fundamental of memristor in Section II. And architecture of the proposed multiplier is presented in Section III. With the multiplier, performance evaluation with a few applications which operate multiplications frequently is presented in Section IV. The evaluation is done with memristor SPICE model and proprietary 180-nm CMOS process. Finally, the conclusion is formed in Section V.

## II. FUNDAMENTAL OF MEMRISTOR

An electric circuit could be interpreted with 4 circuit elements: current ($i$), voltage ($v$), charge ($q$), and magnetic flux ($\Phi$). Six correlations are defined by choosing every 2 elements in total 4 elements. The constitutional formulae are expressed with (1) - (3). Furthermore, the relationship between charge and current is given in (4) and the relationship between magnetic flux and voltage is formulated as (5). Therefore, the relationship between magnetic flux and charge cannot be expressed with the 3 existing elements: resistor, capacitor, and inductor.

$$v = Ri \quad (1)$$

$$q = Cv \quad (2)$$

$$\Phi = Li \quad (3)$$

$$q(t) = \int_{-\infty}^{t} i(\tau)d\tau \quad (4)$$

$$\Phi(t) = \int_{-\infty}^{t} v(\tau)d\tau \quad (5)$$

Memristor has 2 terminals and could serve as a memory and logical device. In 1971, Chua predicted the existence of memristor defined as the connection between magnetic flux and charge [15]. Then the formula is shown in (6); sign M is a memristance constant. Consequentially memristor is defined as the 4th element in electric field.

$$d\Phi = Mdq \quad (6)$$

Based on above theory, researchers with Hewlett-Packard implemented memristor as a real device in 2008 [11]. They realized a memristor which has a structure that $TiO_2/TiO_{2-x}$ insulator layers are combined between platinum. The layers have a slight difference in oxygen combination concentration. On the basis of the structure, characteristics of memristor are taken by ionic migration through external bias. Hereafter, many researchers for increasing performance with maintaining characteristics have been proposed.

Memristor could be operated variously with digital signal processing and it has many application fields. Recently, paper about nano-scale memristor with GHz speed is published in [16]. It is expected the speed of memristive device will exceed the speed of flash memory. Beyond that, the memristor is expected to be a function as synapse of neural computing [17, 18].

## III. RECONFIGURABLE MULTIPLIER ARCHITECTURE BASED ON MEMRISTOR-CMOS TECHNOLOGY

The memristor-CMOS reconfigurable multiplier provides flexible configuration for implementation of multiple multiplier array in real-time. The distinction of the proposed architecture is that designer can choose bit-width of input data by utilizing a horizontal and a vertical control signals and achieve high flexibility. The multiplier can be configured to any bit-width depending on the requirements of a target application. Fig. 1 shows a block diagram for the 1-bit reconfigurable multiplier primitive. CTRLV and CTRLH are the horizontal and the vertical control signals. The exclusive-OR gate decides to propagate input data A and B to the full adder. When the control signals are different, input data are able to be propagated to the full adder. Block that could not bring input data from the exclusive-OR gate would receive only carry and sum from previous block through SIN of sum-in and CIN of carry-in.

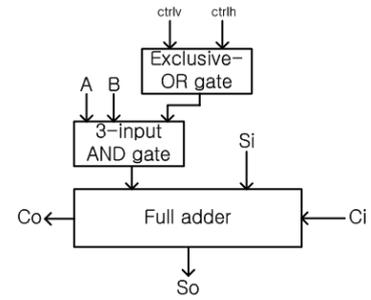

Fig. 1. Block diagram of a 1-bit structure of proposed multiplier. An exclusive-OR gate is used to enable or disable input data, A and B, to be propagated to full adder with control signals, CTRLV and CTRLH.

It has been shown that memristors operating compositely can be used to generate nonlinear dynamics via coupling [19-20], and as logical computational blocks. Memristor-CMOS logic is principally targeted into NAND and NOR gates as shown in Fig. 2 (a) and (b). The threshold logic level at either of the nodes 'x' of the figure is determined by Kirchhoff's current law with $RC_g$ time constant. $R$ is the equivalent parallel resistance of the two memristors depending on its memristive state and $C_g$ is the gate capacitance of the inverter.

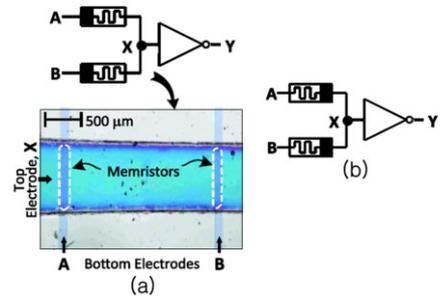

Fig. 2. Block diagram of a memristor-CMOS based NAND and NOR gates : (a) NAND gate, (b) NOR gate.

The approach can be extended to an array architecture having variable bit-width. 8-bit reconfigurable multiplier which consists of 64-block is shown in Fig. 3. The circuit takes the form of ripple carry adder. Each of the blocks in the figure consist of an exclusive-OR gate, a 3-input AND gate, and a full adder introduced in Fig. 1. Multiplier with ripple carry adder form generates partial products in the form of column, and adds the partial products to build output products. Though the structure has more partial products than other structures such as structures with carry save adder form, it has the advantages of that area occupation is smaller and wiring is easier.

Fig. 4 illustrates parallel multiplication of 5-bit and 3-bit on an 8-bit multiplier. Enabled and disabled blocks are decided by control signals. Blocks shown in gray are activated and for 5-bit multiplication. Blocks shown in black are also activated and for 3-bit multiplication. In these multiplications, the horizontal control signals are '11100000' and the vertical control signals are '00011111'. If each of the parts is not overlapped, independent and parallel multiplication will be computed accurately.

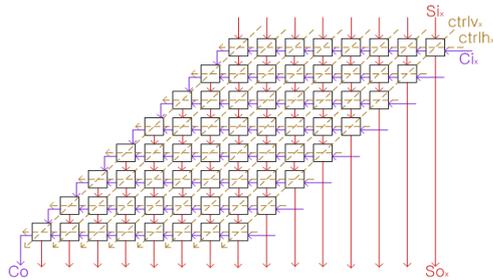

Fig. 3. Block diagram of a proposed 8-bit multiplier. The multiplier takes the form of ripple carry adder. Each of the blocks consist of a exclusive-OR gate, 3-input AND gate and a full adder introduced in Fig. 1.

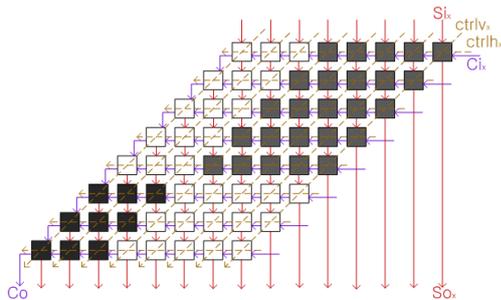

Fig. 4. Block diagram of a 5-bit and 3-bit parallel multiplication on an 8-bit proposed multiplier. Blocks shown in gray are activated for 5-bit multiplication. Blocks shown in black are activated for 3-bit multiplication. Other blocks in white are deactivated.

Proposed multiplier has to secure maximum bit-width of output data depending on bit-width of input data, when reconfiguring the circuit by control signals. For example, when each of input data has 2-bit in bit-width, maximum bit-width of output data is 4-bit. In order to export accurate 4-bit output, the blocks which do not operate are used for data propagation tunnel. The non-operating blocks have no or fewer input data and power dissipation is becoming decreased.

Data propagation process of the proposed multiplier is introduced in Fig. 5. The figure presents functionality of each of blocks when doing a 4-bit multiplication on an 8-bit multiplier. Blocks shown in gray are turned on by control signals and receive 4-bit input data. Blocks shown in black are turned off by control signals and used for data propagation tunnel. Furthermore, blocks shown in white are also turned off and unused for the multiplication.

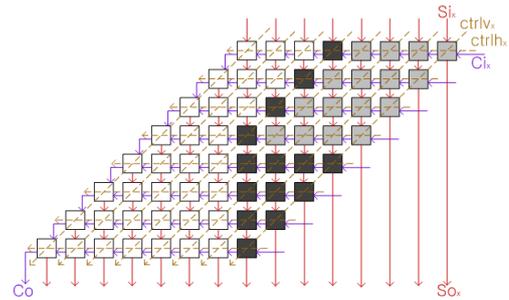

Fig. 5. Block diagram of data propagation process of a 4-bit multiplication on a proposed 8-bit multiplier. Blocks shown in gray are turned on by control signals and receive 4-bit input data. And blocks shown in black are turned off by control signals and used for data propagation tunnel.

Multipliable area on proposed multiplier is decided by control signals. The area which is split into 2-bit parallel and split into 3-bit and 1-bit is introduced in Fig. 6. Partial products are displayed with 2-bit numbers of control signals' status. High status is shown in gray and low status is shown in white. The multiplier is able to compute partial multiplication by way of control signals, irrespective of bit-width.

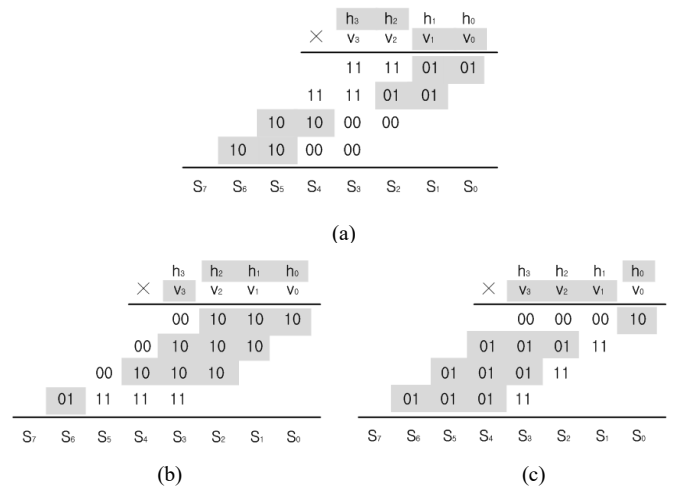

Fig. 6. Illustration of multipliable area on a proposed 4-bit multiplier. The area is decided by control signals. Activated area is shown in gray : (a) 2-bit parallel multiplications, (b) - (c) 3-bit and 1-bit multiplications.

## IV. PERFORMANCE EVALUATION WITH APPLICATIONS

This paper evaluated performance of the proposed multiplier with some of conventional multipliers and through a few applications. The evaluation is done with memristor SPICE model and proprietary 180-nm CMOS process.

Performance comparison with conventional multipliers is introduced in Table I [21]. The proposed multiplier has similar performance with conventional multipliers. However, improved flexibility comparing with conventional multipliers is an advantage of the proposed multiplier.

Block diagram of 4-tap Finite Impulse Response (FIR) filter which is utilized for performance estimation is shown in Fig. 7. The FIR filter consists of 3 types of blocks: D Flip-Flop (DFF), MULT, and ADD. Input and output bit-width of each of the blocks is shown with symbol "[ ]". Frequency of input is 100-MHz.

Overall structure of 4-point parallel input Fast Fourier Transform (FFT) is shown in Fig. 8. (a) [22]. Additionally, blocks of MULTx and RADIXx are presented in Fig. 8. (b) and (c). All of input and twiddle factor, W, have 8-bit input and 100-MHz of frequency. Finally, result of performance test with FIR filter and FFT is introduced in Table II.

Implementation of FIR filter and FFT using the proposed technique reduces about 30% in delay and 35% - 49% in power consumption. Applications which have bigger in bit-width and operate multiplication more frequently will be implemented more efficiently with the proposed technique. Furthermore, designer-selectable bit-width is the heart of the proposed multiplier.

TABLE I. PERFORMANCE COMPARISON OF PROPOSED MULTIPLIER WITH CONVENTIONAL MULTIPLIERS

|  | CMOS | | | Memristor-CMOS |
|---|---|---|---|---|
|  | Conventional RCA | Twin-precision | Scalable | Proposed |
| Bit-width [bit] | 32 | 32 | 32 | 32 |
| Delay [ns] | 11.8 (1.00) | 11.9 (1.01) | 17.3 (1.47) | 12.5 (1.06) |
| Average Power [mW] | 10.9 (1.00) | 11 (1.01) | 18 (1.65) | 11.2 (1.03) |
| Area [$k\mu m^2$] | 61.7 (1.00) | 65.1 (1.06) | 130 (2.11) | 51.2 (0.83) |

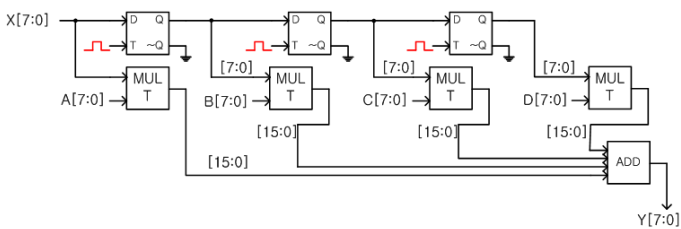

Fig. 7. Block diagram of a 4-tap Finite Impulse Response (FIR) filter circuit. Block MULT becomes 8-bit multiplier and ADD becomes 16-bit adder. Truncation is looked at both MULT and ADD.

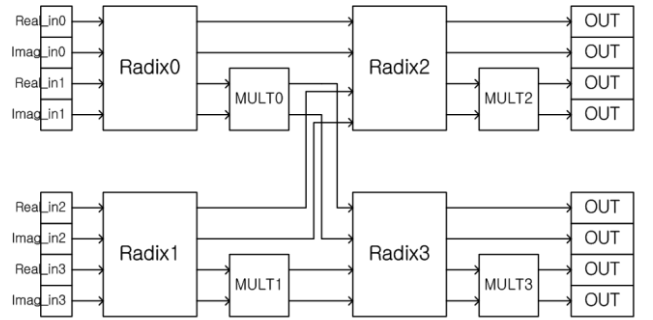

(a)

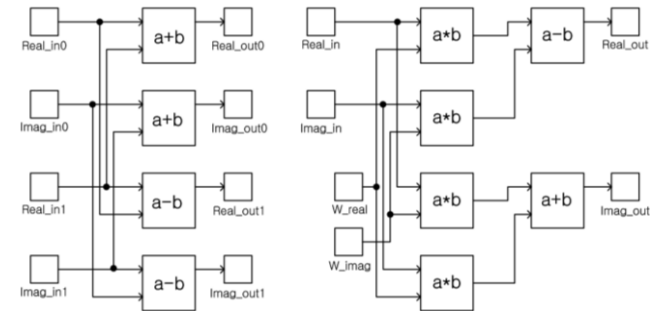

(b)  (c)

Fig. 8. Block diagram of a 4-point parallel input Fast Fourier Transform (FFT) circuit : (a) Overall FFT structure, (b) – (c) Block Radix and MULT. Twiddle factor, W, is random number in 8-bit on the simulation.

TABLE II. PERFORMANCE EVALUATION OF PROPOSED MULTIPLIER THROUGH FIR FILTER AND FFT

|  | Memristor-CMOS proposed | | | |
|---|---|---|---|---|
|  | FIR filter | | FFT | |
| Bit-width [bit] | 8 | 4+4 | 8 | 4+4 |
| Delay [ns] | 28.8 (1.00) | 20.3 (0.70) | 24.5 (1.00) | 16.17 (0.66) |
| Average Power [mW] | 55.8 (1.00) | 36.27 (0.65) | 112.9 (1.00) | 57.58 (0.51) |

V. CONCLUSION

This paper presents memristor-CMOS based reconfigurable multiplier which provides flexible multiplication according to various bit-width. Performance of proposed multiplier is estimated with some applications and comparison with conventional multipliers. The proposed structure could operate diverse parallel multiplication, and reduce power dissipation and computation time.